\medskip

\documentstyle[11pt,epsf]{article}
\begin{document}
\setlength{\baselineskip}{20pt}

\def\al{\alpha}
\def\be{\beta}
\def\ga{\gamma}
\def\de{\delta}
\def\vep{\varepsilon}
\def\th{\theta}
\def\ka{\kappa}
\def\si{\sigma}
\def\ph{\phi}
\def\om{\omega}
\def\Ga{\Gamma}
\def\Ps{\Psi}
\def\Si{\Sigma}
\def\ov{\over}
\def\pd{\partial}

\centerline{ {\bf  On the Probability Distribution of
Velocity Circulation} }
\centerline{ {\bf  in Three-Dimensional Turbulence } }
\medskip

\bigskip

\centerline{ Makoto Umeki }

\centerline{Department of Physics, University of Tokyo,
Hongo, Bunkyo-ku, Tokyo 113, Japan}
\centerline{Email: umeki@tansei.cc.u-tokyo.ac.jp}
\centerline{ (Received, \hskip 5cm \ \ ) }

\bigskip

\centerline{{\bf ABSTRACT}}

The probability distribution functions of
the circulation of velocity in three-dimensional decaying
isotropic turbulence are examined by the database of
the numerical simulation based on the pseudospectral method.
It is shown that the standard deviation increases nearly
as $A^{2/3}$ where $A$ is the area of the loop.
The PDFs change from exponential to gaussian as the
size becomes large.
The former agrees well with the prediction by Migdal (1993),
although the latter does not match.
A modification is proposed to explain the gaussian distribution
with the $A^{2/3}$ dependence.

\bigskip


\bigskip

The characteristic functional equation of fluid turbulence
initiated by Hopf remains unsolved in spite of its elegant
and rigorous formalism. Recently, Migdal (1993) applied
the loop calculus developed in the quantum field theory
to the Hopf functional equation of
three-dimensional Navier-Stokes turbulence.
The Hopf functional is defined by the loop functional as
$$ \Ps (C) = \Bigl< \exp \Bigl[ {i \over \nu}
\oint_C dr_\al v_\al (r) ) \Bigr] \Bigr>, \eqno(1)$$
where the loop is parametrized by the angular variable
$\th \in [0, 2\pi ]$ and $\nu$ is the kinematic viscosity of the
fluid. The Navier-Stokes equation integrated along the loop
is rewritten by
$$i\nu \dot{\Ps} [C] = H_C \Ps [C] ,  \eqno(2)$$
$$H_C \equiv \nu^2 \oint_C dr_\al
\Bigl[i\pd _\be {\de \over \de \si_{\be \al}(r) }
+ \int d^3 r' { r'_\ga - r_\ga \over 4\pi |r-r'|^3}
{\de^2 \ov \de \si_{\be \al}(r) \de \si_{\be \ga}(r') } \Bigr] ,
\eqno(3)$$
Here the {\it area derivative} $\de / \de \si_{\mu \nu}(r)$ of the
loop functional is related to the vorticity as
$${i \nu \de \Ps \ov \de \si_{\nu \mu}(r) } =
< \om_{\mu \nu} (r)>. \eqno(4)$$
Regarding the inviscid limit $\nu \rightarrow 0$ as the WKB approximation,
the functional will be represented by the action $S$ as
$$\Ps [C] = \exp (iS[C] / \nu ).                            \eqno(5)$$
Neglecting the viscous and forcing terms,
the functional equation for $S$ becomes
$$\dot{S}[C] = \oint_C dr_\al \int d^3 r'
{ r'_\ga - r_\ga \ov 4\pi |r-r'|^3}
{ \de S \ov \de \si_{\be \al} (r)}
{ \de S \ov \de \si_{\be \ga} (r')} .                        \eqno(6)$$
Equation (6) possesses scaling solutions
$S[C]= t^{2\ka -1} \ph [C/t^{\ka}] $ with arbitrary $\ka$.
In order to find the turbulence solution, the length scale should
be constructed by the energy dissipation rate $\vep$.
Thus we obtain the action
$$S[C] = \vep t^2 \ph \Bigl[{C \ov \sqrt{\vep t^3} } \Bigr] , \eqno(7)$$
with the Kolmogorov index $\ka = 3/2$.
Assuming the stationariness by taking
$\ph (x) = iB x^{4/3}$, where $B$ is a positive universal constant,
and replacing $C^2$ by the area $A$ of the loop,
the action becomes
$$S[C] = i B \vep^{1/3} A^{2/3} .                          \eqno(8)$$
At the final step, Migdal made two assumptions in (5) to
derive the probability
distribution of the velocity circulation; the first is the
introduction of the parameter $\ga$ as $|\ga|$ and the second is
the choice of $A$ as tensor area
$A=\Si^C_{\al \be}= \oint_Cr_\al dr_\be$.
Then the functional becomes
$$\Ps [C] = \exp \Bigl[-B|\ga |
({\vep \ov \nu^3} (\Si_{\al \be} ^C)^2)^{1/3} \Bigl].       \eqno(9)$$
By Fourier transformation, the lorentzian distribution is obtained as
$$P(\Ga )= {\Ga_w \ov \pi (\Ga^2 + \Ga_w^2) },             \eqno(10)$$
where the width is expressed as
$\Ga_w =B \vep^{1/3} (\Si_{\al \be}^C)^{2/3}$ .
These predictions are worthy to be examined by both numerical and
experimental data. However, it should be noted that
the lorentzian distribution has an infinite variance which may not
be observed exactly. In this paper we compare them with a numerical
simulation because the comparison with the experimental data
seems very difficult.

The predictions by Migdal are examined by the database of
the pseudospectral simulations with resolution $128^3$
performed by Yamamoto \& Hosokawa (1988), in which the viscosity
$\nu$ is taken as $0.002$ and the Reynolds number based
on the Taylor's microscale is about 100. The data at $t=10$
is used.

We choose a square of length $L$ as the loop and move it through all
possible positions. By changing $L$ from $d$ to 16$d$, where $d$
is the size of the mesh, the dependence of the PDFs of the velocity
circulation and its width on the area of the loop is obtained.
The inertial range can be estimated as the scale from 4$d$ to 16$d$.

Figure 1 shows the PDF normalized by the numerically obtained
standard deviation. When the size is in the viscous scale $L \le 4d$,
the PDF is nearly exponential. As the size becomes larger,
the PDF changes gaussian. Thus we do not observe the appearance of
the lorentzian distribution.

Figure 2 shows the dependence of the standard deviation $\Ga_s$
and the most probable width $\Ga_w$ obtained by the least square method
on the size of the loop.
Here the integration range in the fitting is taken
from the minimum to the maximum of the numerical data.
If the vorticity at each point inside the loop were independent,
the standard deviation would scale sa $A^{1/2}$.
It is observed that both $\Ga_s$ and $\Ga_w$ are nearly
proportional to $A^{2/3}$.
The least-square fitting leads that the power of $A$ is 0.5996 for
$\Ga_s$ and 0.6740 for $\Ga_w$.

There is a question on the hypothesis used by Migdal that
the action depends on the {\it tensor} area.
For the Wilson loop in QCD the action is a function of the
{\it scalar} area. Here it is appropriate to define the scalar
area as the minimal area of the surface enclosed by the loop.
In order to examine this hypothesis, the loop is taken as
an eight-figure shape made by two squares of size $L_1$ and
$L_2$ with $L_1=L_2+1$, which have parallel sides and are
connected at a single common vertex. The direction of integration
is taken as opposite in the two squares. $L_2$ is varied from 1 to 11.
If the tensor-area law is correct,
the standard deviation should scale as $(L_1^2-L_2^2)^{2/3}
= (2L_2+1)^{2/3}\approx L_1^{2/3}$.
Otherwise, the scalar-area law will lead $(L_1^2+L_2^2)^{2/3} \approx
L_1^{4/3}$.
Figure 3 shows the dependence of $\Ga_s$ and $\Ga_w$ on the scalar
area.
It is shown that both $\Ga_s$ and $\Ga_w$ scale nearly as
$(L_1^2+L_2^2)^{2/3}$.
The fitting leads that the power is 0.6357 for
$\Ga_s$ and 0.7277 for $\Ga_w$.
According to this numerical analysis, the hypothesis of the area law
should be modified to the scalar-area law.

The dependence is also examined of the PDFs on the shape of the
rectangular loop by changing the aspect ratio.
The area of the rectangle is fixed as 240$d^2$ and the sides are
changed as $(L_x, L_y)=(l_xd, l_yd)$, where $(l_x, l_y)$ is the
pair of integers whose product is 240, i.e. (2, 120), $\cdots$
(15,16). The PDFs are almost gaussian in all cases.
The dependence of the standard deviation and the width on
the total length of the sides is shown in Figure 4.
It is found that the variance is little affected by the
total length. Thus the scalar-area law is also supported
by this numerical result.

It should be noted that the scaling solution (8) of the action $S$
may be obtained by the dimensional analysis, i.e. $S$ should be
a function of only the energy dissipation rate and the area of the
loop.
We find it possible to match the Migdal's original theory
with the numerical results. First, the parameter
$\ga$ should be introduced quadratically in (9).
It will lead to the gaussian probability distribution.
It may be justified by considering the circulation is
the area integral of the vorticity
$$\Ga[C] = \int_S \mbox{rot \boldmath $v$} dS, \eqno(11)$$
because the area integral scales $\ga^2$ when we regard $\ga$ as
the length of the loop.
Second, as we see in the numerical analysis,
the replacing $C^2$ in (8) by the scalar area $A$.

This first investigation of the PDF of the velocity circulation
should be confirmed by other numerical simulations with the higher
resolution, e.g. 256$^3$ or 512$^3$.

\bigskip

{\bf \S Acknowledgement}

I would like to acknowledge Professor K. Yamamoto for providing the
database of the numerical simulation of the three-dimensional
turbulence and Dr. Y. Matsuo and Dr. H. Cateau for useful discussions.

\bigskip

{\bf \S References}

Migdal, A. A. 1993 Loop equation in turbulence.
PUTP-1383. preprint

Yamamoto, K., \& Hosokawa, I. 1988
A decaying isotropic turbulence pursued by the spectral method.
{\it J. Phys. Soc. Jpn.}  {\bf 57} 1532-1535.

\bigskip

{\bf \S Figure Captions}

Figure 1. Normalized PDFs of the velocity circulation $\Ga$
versus the area of the square loop.
Four cases $L/d= 2$ (solid line), 4 (dased), 8(dotted), 16(dotdashed)
are drawn.
Thin solid and dashed lines
denote the gaussian and the exponential distributions
respectively.

Figure 2. The dependence of the standard deviation (denoted by circles)
and the numerically fitted width (squares) on the area of the
square loop. Solid and dashed lines denote $A^{1/2}$ and $A^{2/3}$
respectively.

Figure 3. The dependence of the standard deviation
(denoted by circles)
and the numerically fitted width (squares) on the scalar area
of the eight-shaped loop. Solid and dashed lines
denote $A^{1/2}$ and $A^{2/3}$ respectively.

Figure 4. The dependence of the standard deviation
(denoted by circles)
and the numerically fitted width (squares) on the total length of
the sides of the square loop.

\begin{center}
\epsfile{file=fig1.ps}
\end{center}

\begin{center}
\epsfile{file=fig2.ps}
\end{center}

\begin{center}
\epsfile{file=fig3.ps}
\end{center}

\begin{center}
\epsfile{file=fig4.ps}
\end{center}

\end{document}